\documentstyle[prd,epsfig,aps,preprint]{revtex}
\begin{document}

\draft
\title{The Existence of an Old Quasar at $z = 3.91$ and its
Implications for $\Lambda(t)$ Deflationary Cosmologies}

   \author{J. V. Cunha\footnote{jvital@dfte.ufrn.br} and R.
   C. Santos\footnote{rose@dfte.ufrn.br}
          }

  \smallskip
\address{~\\Departamento de F\'{\i}sica, \\Universidade Federal
             do Rio Grande do Norte, C.P. 1641, 59072-970, Natal,
             RN, Brasil
                  }

\date{\today}
\maketitle


\begin{abstract}We investigate some observational constraints on decaying vacuum
cosmologies based on the recently discovered old high redshift
quasar APM 08279+5255. This object is located at $z = 3.91$ and
has an estimated age of 2-3 Gyr. The class of $\Lambda(t)$
cosmologies is characterized by a positive $\beta$ parameter
smaller than unity which quantifies the ratio between the vacuum
and the total energy density. Assuming the lower limit age (2 Gyr)
and that the cold dark matter contributes with $\Omega_{\rm
M}=0.2$ we show that $\beta$ is constrained to be $\ge 0.07$ while
for an age of 3 Gyr and $\Omega_{\rm M}=0.4$ the $\beta$ parameter
must be greater than $0.32$. Our analysis includes closed, flat
and hyperbolic scenarios, and it strongly suggests that there is
no age crisis for this kind of $\Lambda(t)$ cosmologies. Lower
limits to the redshift quasar formation are also briefly discussed
to the flat case. For $\Omega_{\rm M}=0.4$ we found that the
redshift formation is constrained by $z_{f}\ge 8.0$.
\end{abstract}


\section{Introduction}

Recent observations from Supernovae (SNe) type Ia  strongly
suggest that the bulk of energy in the Universe is repulsive and
appears like a dark component; an unknown form of energy with
negative pressure [in addition to the ordinary dark matter] which
is probably of primordial origin\cite{PR}. The most natural
candidate for dark energy is the cosmological constant
($\Lambda$), or equivalently, a perfect fluid obeying the equation
of state, $p_v = - \rho_v$, which is usually interpreted as the
constant vacuum energy density of all fields existing in the
Universe. The $\Lambda$-term is the simplest but not the unique
possibility. Other candidates appearing in the literature are: a
relic scalar field component (SFC) which is slowly rolling down
its potencial\cite{ratra}, a decaying vacuum energy density, or a
time varying $\Lambda$-term\cite{OT}, the so-called ``X-matter",
an extra component\cite{turner} characterized by an equation of
state $p_{\rm x}=\omega\rho_{\rm x}$ (XCDM), and the Chaplygin
type gas whose equation of state is $p= -A/\rho^{\alpha}$, where
$A$ and $\alpha$ are positive constants\cite{bert} (see also
Lima\cite{LIMAR} for a quick review). On the other hand, the
existence of old high-redshift objects is one of the best methods
for constraining the age of the Universe, as well the basic
cosmological parameters\cite{Krauss97}. Such objects also provide
an important key for determining the first epoch of galaxy
formation. In this connection, quasars are among the most luminous
objects known in the universe and their prominent emission-lines
contains valuable information to estimate their ages. The recently
reported age estimates of the APM 08279+5255 quasar with a lower
limit of 2-Gyr-old at redshift $z=3.91$ is therefore a
particularly interesting event\cite{KOM}. In this article we
investigate some cosmological implications from the existence of
this quasar to a large class of decaying vacuum cosmologies
proposed by Lima and collaborators\cite{LM}. Some constraints on
the first epoch of quasar formation are also discussed.

\section{Old High-z Objects and Age-redshift Test}

Several groups have searched for old objects at high
redshifts\cite{old}. Such discoveries accentuated even further the
already classical  ``age crisis" and gave rise to a new variant of
this problem, which could be named the high-$z$ time scale
crisis\cite{LA}. More recently, Hasinger and co-workers \cite{KOM}
reported the discovery of an ionized Fe K edge in the $z=3.91$
broad absorption line quasar APM 08279+5255. The Fe/O ratio of the
absorption material is significantly higher than solar abundance
(Fe/O$=2-5$), and requires a timescale of $\sim 3$ Gyr with a
conservative lower limit age of $2$ Gyr. The Einstein-de Sitter
model is not able to explains the age of this
object\cite{KOM,ALV}. Let us now examine how the existence of this
quasar constrains the basic parameters defining the time varying
$\Lambda(t)$ cosmologies. The general age-redshift relation in the
$\Lambda(t)$ models proposed by Lima and coworkers\cite{LM} is
\begin{eqnarray}
t_z & = &H_o^{-1} \int^{(1 + z)^{-1}}_{0} {dx \over \left[1 -
\frac{\Omega_{\rm M}}{(1 - \beta)} + \frac{\Omega_{\rm M}}{(1 -
\beta)} x^{3\beta - 1}\right]^{{1}\over{2}}}\equiv
H_{o}^{-1}f(\Omega_{\rm M}, \beta, z) .
\end{eqnarray}
For $\beta=0$ and $\Omega_{M}=1.0$ the above expression reduces to
$t_z = \frac{2}{3}H_o^{-1}(1+z)^{-3/2}$, which is the prediction
of the Einstein-de Sitter model\cite{KT}. In this case, the age of
the Universe at redshift $3.91$ is nearly 1 Gyr which is much less
than the lower limit of the quasar age. Following standard lines,
we take for granted that the age of the Universe at the specific
redshift is bigger than or at least equal to the age of the
quasar, namely:
\begin{equation}
\frac{t_z}{t_q} = \frac{f(\Omega_{\rm M}, \beta, z)}{H_o t_q} \geq
1,
\end{equation}

where $t_q$ is the estimated age of the quasar and $f(\Omega_{\rm
M},\beta, z)$ is the dimensionless factor defined by  Eq. (1).
Note that the denominator of the above expression defines a
dimensionless age parameter $T_q = H_o t_q$. For the lower limit
age of the quasar its value and the most recent determinations of
the Hubble parameter ($H_o = 72 \pm 8$ ${\rm{km s^{-1}
Mpc^{-1}}}$)\cite{F01}, such a quantity takes values on the
interval $0.131 \leq T_q \leq 0.163$ so that $T_q \geq 0.131$.
Therefore, for a given value of $H_o$, only models having an
expanding age bigger than this value at $z = 3.91$ will be
compatible with the existence of this object. In order to assure
the robustness of our analysis, we will adopt in our computations
the lower bound for the above mentioned value of the Hubble
parameter, i.e., $H_o = 64$ ${\rm{km s^{-1} Mpc^{-1}}}$.

The figures above show the dimensionless age parameter $T_z = H_o
t_z$ (see Eq. (1)) as a function of the redshift for several
values of $\beta$ with $\Omega_{\rm M}=0.2$ (a, b), and
$\Omega_{\rm M}=0.4$ (c,d). The shadowed regions in the graphs
were determined from the minimal value of $T_q$. Models with
curves crossing the rectangles are ruled out because they yield an
age parameter smaller than the minimal value required by the
presence of the quasar APM 08279+5255. By assuming an age estimate
of 2 Gyr one may see from plot (a) that the minimal value for the
vacuum energy density is $\beta \geq 0.07$, which provides a
minimal total age of $\sim 13.4$ Gyr. From Figure (b), we see that
the minimal value of $\beta$ required is $0.31$. For $\Omega_{\rm
M}=0.4$, the minimal values of $\beta$ are 0.15 and 0.32 for 2 and
3 Gyr, respectively (see also table arising on Figure 2). Such
limits are less stringents than the values inferred from angular
size-redshift relation where $\beta$ falls on the
interval\cite{Cunha} [0.58,0.76]. However, we stress that
constraints from angular size redshift relation should be taken
with some caution because a statistical analysis describing the
intrinsic length distribution of the sources is still lacking.
Further, although being relatively easy for decaying vacuum
cosmologies to solve the age-redshift problem, the same does not
happens for other classes of dark energy models like constant
$\Lambda$ or XCDM cosmologies\cite{LA,ALV}.

\section{Implications on the Epoch of Quasar Formation}

In the present section, we discuss how the age estimates of the
quasar APM 08279+5255 constrain the epoch of quasar formation.
Since we have neglected a possible incubation time in our
computations, the obtained redshift formation $z_f$ will be a
conservative lower bound. For look-back time calculations, such a
hypothesis can be translated as\cite{LA}
\begin{eqnarray}
t_{obs}-t_{z_f}&=&H_o^{-1}\int\limits_{1 \over {1 + z_f}}^{1 \over
{1 + z_{obs}}}{dx \over \left[1 - \frac{\Omega_{\rm M}}{(1 -
\beta)} + \frac{\Omega_{\rm M}}{(1 - \beta)} x^{3\beta -
1}\right]^{{1}\over{2}}} \geq t_q,
\end{eqnarray}
where the inequality signal comes from the fact that the Universe
is older than or at least has the same age of any observed
structure. Naturally, models for which $z_f \rightarrow \infty$
are clearly incompatible with the existence of this particular
object, being ruled out in a natural way. In Fig. 2 we show  the
$z_f - \Omega_{\rm M}$ plane allowed by the existence of the
quasar APM 08279+5255 for $\Lambda(t)$ models. As before, two age
estimates were assumed: 2 Gyr (solid curve) and 3 Gyr (dashed
curve). As should be physically expected, since the effect of dark
matter is decelerate the cosmic expansion\footnote{It means that
the look-back time between the observed redshift $z_{obs}$ and
$z_f$ is smaller for larger values of $\Omega_{\rm M}$.}, the
larger the contribution of $\Omega_{\rm M}$ the larger the value
of $z_f$ that is required in order to account for the existence of
this quasar within these cosmological scenarios.

In this way, the smallest value for the formation redshift occurs
for a completely empty universe ($\Omega_{\rm M} = 0$). We find
$z_f \geq 5$. For a low-density universe with $\Omega_{\rm M} =
0.2$ we obtain $z_f \geq 5.85$. For an universe with $\Omega_{\rm
M} = 0.4$ we obtain $z_f \geq 8.04$. The basic results are
presented in the Table appearing in Figure 2.

Summarizing, we have investigated some observational constraints
on deflationary cosmologies provided by the age of the old quasar
APM 08279+5255. If the estimated age of this object is correct, at
light of the above results, we conclude that such models are more
efficient to explain the old quasar at $z=3.91$ than constant
$\Lambda$ dark energy models. Even for a value of $\beta \ge 0.07$
the age crisis at high $z$ is solved. Concerning the first epoch
of quasar formation the minimal redshift formation required is
$z_{f}\geq 8.0$. Naturally, new tests and analysis must be
developed in order to know if deflationary cosmologies may be
realistic models of the Universe.

{\bf Acknowledgements:}The authors are grateful to Prof. J.A.S.
Lima for helpful discussions. This work was supported by CAPES
(Brazilian Research Agency).

\begin{figure}
\vspace{.2in}
\centerline{\psfig{figure=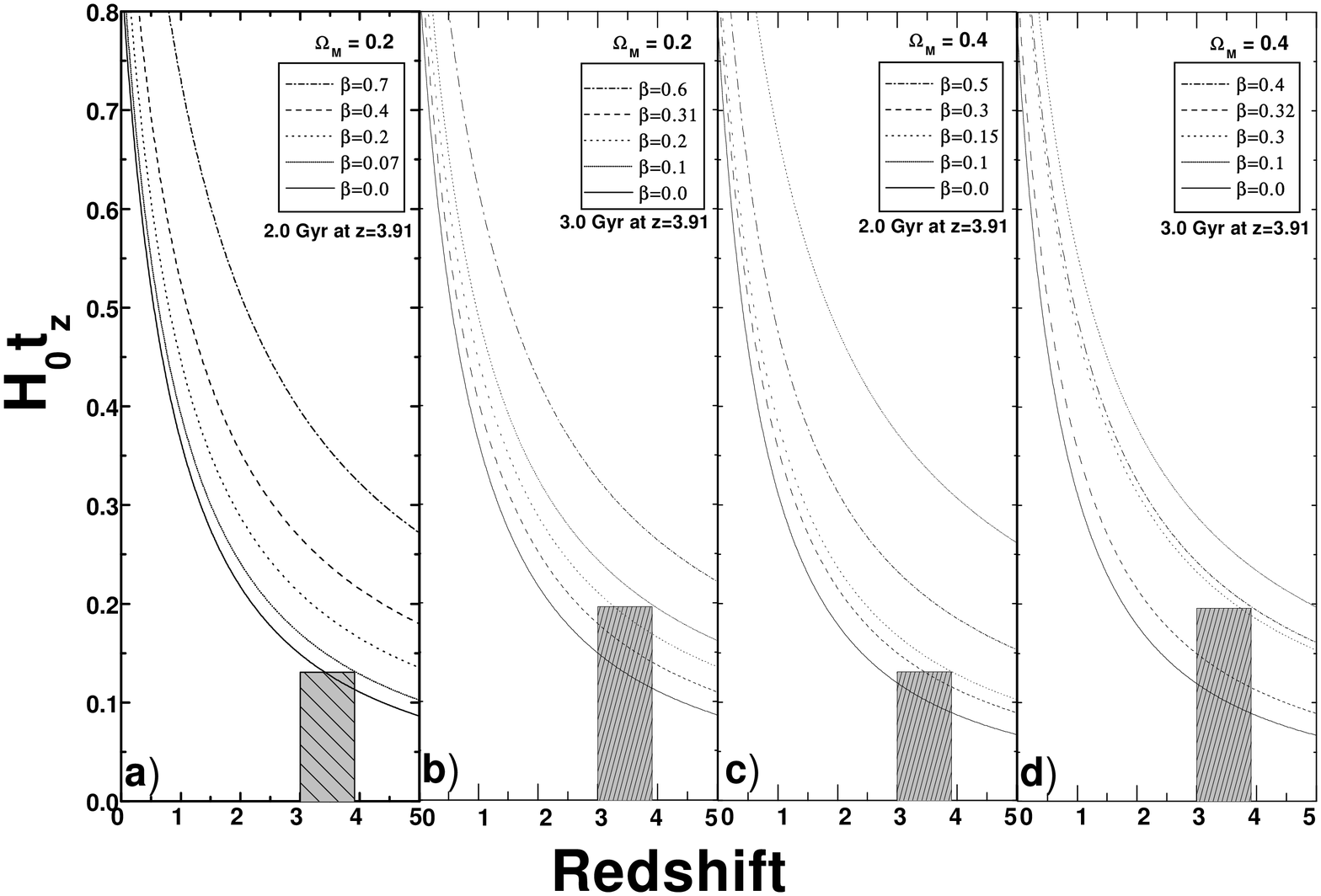,width=5.4truein,height=2.2truein}
}\caption{Dimensionless age parameter as a function of redshift
for some values of $\beta$ with $\Omega_{\rm M}=0.2$ and
$\Omega_{\rm M}=0.4$. As explained in the text, all curves
crossing the shadowed area yield an age parameter smaller than the
minimal value required by the quasar APM 08279+5255. Note that
figures (b) and (d) are the same as in Panel (a) and (c), but for
an estimate age of 3 Gyr.}
\end{figure}

\begin{figure}
\centerline{\psfig{figure=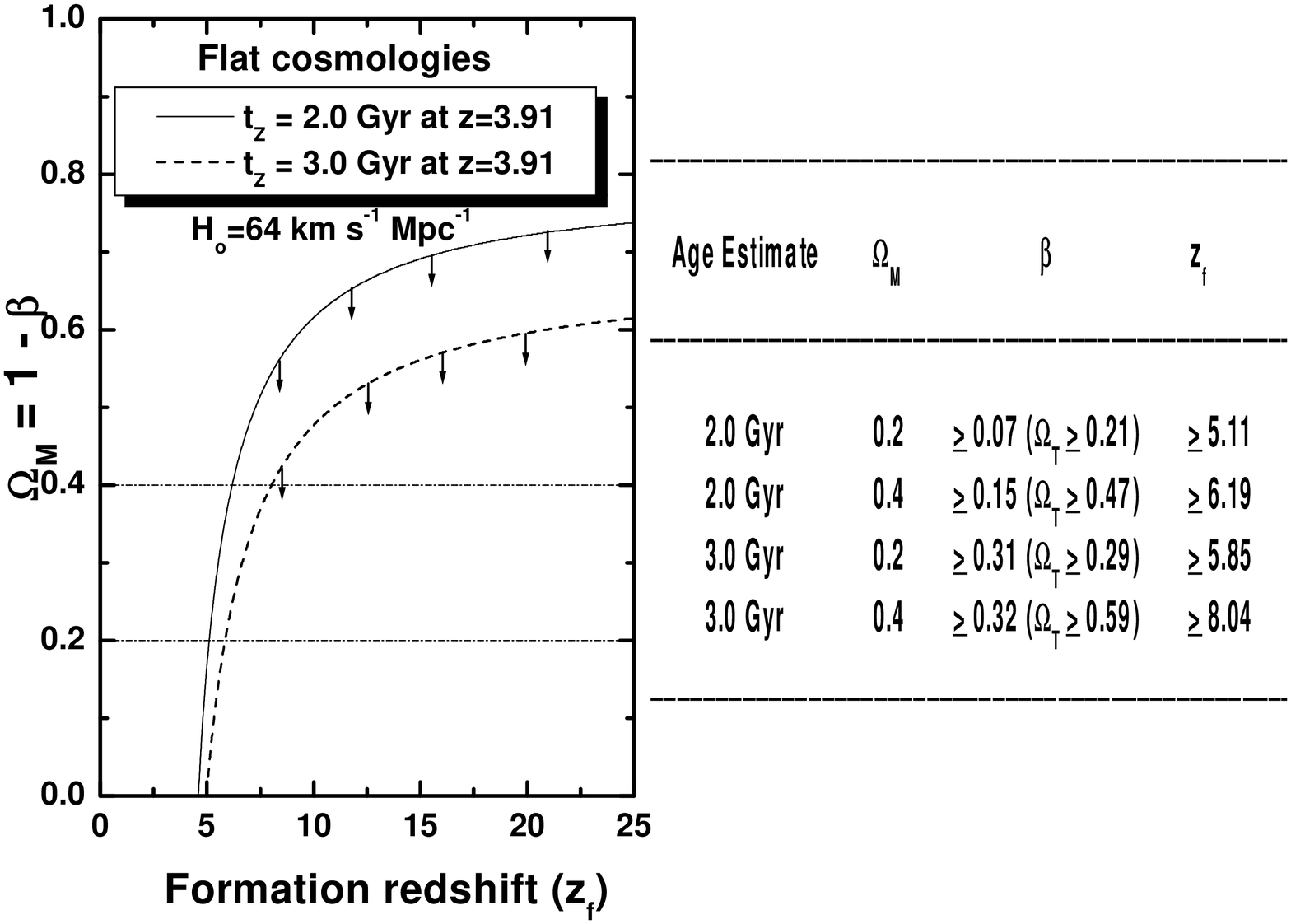,width=5.4truein,height=2.4truein}
\hskip 0.1in} \caption{The formation redshift versus $\Omega_{\rm
M}$ plane for flat $\Lambda(t)$ models. Countours of a fixed age
parameter $H_o t_z$ for the quasar APM 08279+5255. The solid curve
corresponds to an age estimate of 2 Gyr while the dashed one
stands for 3 Gyr. For each contour the arrows point to the allowed
parameter space.}
\end{figure}

\end{document}